\documentclass[preprint,showpacs,preprintnumbers,amsmath,amssymb]{revtex4}
\usepackage{bm}
\usepackage{epsf}

\everymath{\displaystyle}
\begin{document}           
\title{Comparison of Spin Dynamics in the Cylindrical
and Frenet-Serret Coordinate Systems}

\author{Alexander J. Silenko}\email{alsilenko@mail.ru}
\affiliation{Bogoliubov Laboratory of Theoretical Physics, Joint Institute for Nuclear Research, Dubna 141980, Russia,\\
Research Institute for Nuclear Problems, Belarusian State University, Minsk 220030, Belarus}

\begin{abstract}
A comparative analysis of a description of spin dynamics in the cylindrical and Frenet-Serret
coordinate systems is carried out. An equivalence of these two systems is shown. A possibility of efficient use of the cylindrical coordinate 
system for a calculation of spin evolution of particles and nuclei in accelerators and storage rings is caused by an immobility of its coordinate axes relative to stationary detectors.
\end{abstract}


\maketitle

\section{INTRODUCTION}

This paper is devoted to a comparative analysis of
describing spin dynamics in the cylindrical and
Frenet-Serret coordinate systems. The latter is necessary because of the importance of precisely measuring
the evolution of particle and nuclei polarizations in
accelerators and storage rings. The basic equation for
spin motion in electromagnetic fields is a Thomas-Bargmann-Michel-Telegdi (T-BMT) one \cite{T-BMT}, given
in Cartesian coordinates. It is very important to search
for electric dipole moments (EDMs) of particles and
nuclei in the experiments prepared for running at storage rings
\cite{fukuyama}. To calculate the spin dynamics in such
experiments, it is necessary to use a generalization of
the T-BMT equation, which takes into account the
presence of EDM (see Ref. \cite{GBMT} and references therein).

The equations mentioned above determine spin
motion in Cartesian coordinates. However, polarized
beams of particles and nuclei in accelerators and storage rings move along practically closed trajectories. In
accelerator theory, a standard choice for the coordinate system is a Frenet-Serret one (FS), whose axis
orientation is fixed by the particle motion. In this case,
the equation of spin motion describes the motion of a
spin (pseudo)vector with respect to the momentum
vector, i.e., the change in the relative orientation of
these vectors. This fact complicates a description of
the spin dynamics. In the real world, one measures
spin orientation relative to the detectors, whose position is fixed in the cartesian coordinates rather than
with respect to the momentum vector, whose projections onto all three axes of a given reference frame
change. In Ref. \cite{RPJSTAB}, an alternative
description and exact equations describing the motion
of a particle and its spin in the cylindrical coordinate
system were found. However, a detailed analysis of differences in
the descriptions of spin dynamics in both coordinate
systems was not carried out in Ref. \cite{RPJSTAB}. Because the problem of particles and nuclei spin dynamics in accelerators and storage rings is vitally important, such an
analysis should be done anyway.

We use the system of units $c=1$.

\section{COMPARISON OF CYLINDRICAL AND FRENET-SERRET
REFERENCE FRAMES}

In a cylindrical coordinate system, the azimuthal
angle $\phi$ is given by the particle position. Upon changing a particle azimuth by the angle $d\phi$ the horizontal
axes of cylindrical and cartesian coordinate systems
turn with respect to each other by the same angle; i.e.,
the cylindrical coordinate system rotates with respect
to the Cartesian one around the $z$ axis with an instantaneous angular velocity $d\phi/dt$ (see Ref. \cite{RPJSTAB}).

The axes of the FS coordinate system are set by the
particle trajectory. The orts of this reference frame are directed along the tangent to the trajectory (parallel to the velocity and
momentum), in the trajectory plane along the normal
to it (parallel to an acceleration vector), and along the
binormal perpendicular to these two orts. Relative to
the Cartesian coordinate system, an FS one rotates
about all three axes, not only around $z$ axis as a cylindrical reference frame. The motion of particles and
nuclei in the vertical direction is oscillatory in accelerators and storage rings.

The peculiar features of the cylindrical and FS coordinate systems can be quantitatively described by setting an
evolution of a unit vector along the vectors of particle velocity and
momentum, $\bm N=\bm v/v=\bm p/p$. From the Lorentz
equation
\begin{equation}
\frac{d\bm p}{dt}=e\left(\bm E+\bm\beta\times\bm B\right), ~~~
\bm\beta=\frac{\bm v}{c}=\frac{\bm p}{\gamma m},
\label{eq1}\end{equation}
it follows that this vector rotates with an instantaneous
angular velocity \cite{RPJSTAB}
\begin{equation}
\bm\omega=-\frac{e}{\gamma
m}\left(\bm B-\frac{\bm N\times\bm E}{\beta}\right).
\label{eqm}\end{equation}


The additional (with regard to the rotation of a
cylindrical coordinate system) rotation of the FS reference frame relative to the Cartesian one occurs in the
presence of nonzero horizontal components of magnetic and/or quasimagnetic fields $\bm B$ and $\bm N\times\bm E$. The
relationship between an angular velocity of $\bm N$ vector
rotation and relative motion of cylindrical and Cartesian coordinate systems found in Ref. \cite{RPJSTAB} is not trivial. Let
us denote the projection of any vector onto a horizontal plane by the symbol $\|$. Since the vector $\bm N$ is directed
along the tangent to the trajectory, the change in the
particle azimuth in this plane equals an angle between
two horizontal projections of this vector, which can be
denoted as $\bm N_\|$ and $\bm N'_\|$. An infinitesimal angle $d \phi$ characterizing the change in the particle azimuth is given
by the following expression:
$$d\phi=\frac{(\bm N_\|\times\bm N'_\|)\cdot\bm
e_z}{|\bm N_\| |\cdot|\bm N'_\| |}=\frac{(\bm N_\|\times d\bm
N_\|)\cdot\bm e_z}{|\bm N_\| |^2}, $$ where $d\bm N_\|=\bm
N'_\|-\bm N_\|$ is an infinitesimal vector. The
instantaneous rate of the azimuth change is found to
be \cite{RPJSTAB}
\begin{equation}
\dot{\phi}\equiv\frac{d\phi}{dt}=\frac{(\bm N_\|\times\dot{\bm
N}_\|)\cdot\bm e_z}{|\bm N_\| |^2}=\omega_z-o,
\label{eq2}\end{equation} where
\begin{equation}
o=\frac{(\omega_xN_x+\omega_yN_y)N_z}{1-N_z^2}=\frac{(\omega_\rho
N_\rho+\omega_\phi N_\phi)N_z}{1-N_z^2}. \label{eqo}\end{equation}
The components of the $\bm\omega$ vector are fixed by
Eq. (\ref{eqm}).

Equations (\ref{eq2}) and (\ref{eqo}) are exact. This is shown in \cite{RPJSTAB} by considering an example of a particle running the
loop whose normal is deviated from the $z$ axis by some
angle. Taking into account a correction due to quantity $o$ allows an exact description of the particle motion
projected onto a horizontal plane. However, the quantity $o$ is usually rather small. If the horizontal
plane coincides with the plane of unperturbed particle
motion, this quantity can usually be neglected (see analysis in \cite{RPJSTAB}).

\section{EQUATION OF SPIN MOTION}

In all reference frames considered so far, the spin
motion is precessional. The effects due to a spin-tensor interaction (see \cite{PRC,PRC2009,dEDMtensor,PRDexact} and references therein) are
considered in the present paper. Let the particle with
electric and magnetic dipole moments moves in the
electromagnetic field. The equation describing the motion of its spin in the Cartesian coordinate system
looks as follows \cite{GBMT}:
\begin{equation}\begin{array}{c}
\frac{d\bm s}{dt}=\bm\Omega\times\bm s, ~~~ \bm\Omega=\bm\Omega_{MDM}+\bm\Omega_{EDM},
\\
\bm\Omega_{T-BMT}=-\frac{e}{m}\left[\left(a+\frac 1\gamma\right) \bm
B-\frac{a\gamma}{\gamma+1}\bm\beta(\bm\beta\cdot\bm B)
\right.
\\ \left.
-\left(a+\frac{1}{\gamma+1}\right)\left(\bm\beta\times\bm
E\right)\right],
\\
\bm\Omega_{EDM}=-\frac{e\eta}{2m}\left(\bm
E-\frac{\gamma}{\gamma+1}\bm\beta(\bm\beta\cdot\bm
E)+\bm\beta\times\bm B\right),\label{eqspinm}\end{array}\end{equation} where the quantities $\bm\Omega_{MDM}$ and $\bm\Omega_{EDM}$ make contributions of electric and magnetic dipole moments,
respectively. The same look is acquired by the quantum mechanical equations of spin motion for particles
with spins 1/2 \cite{RPJ} and 1 \cite{PRDspin} after transition to the classical limit.

An angular velocity of spin rotation in the cylindrical coordinate system is obtained by subtracting the
quantity $\bm\Omega$ from $\dot{\phi}\bm e_z$. Upon neglecting the correction $o$, it is given by the expression
\begin{eqnarray}
\bm\Omega^{(cyl)}=-\frac{e}{m}\left\{a\bm B-
\frac{a\gamma}{\gamma+1}\bm\beta(\bm\beta\cdot\bm B)\right.\nonumber\\
+\left(\frac{1}{\gamma^2-1}-a\right)\left(\bm\beta\times\bm
E\right)+\frac{1}{\gamma}\left[\bm B_\|
-\frac{1}{\beta^2}\left(\bm\beta\times\bm
E\right)_\|\right]\nonumber\\ \left.+ \frac{\eta}{2}\left(\bm
E-\frac{\gamma}{\gamma+1}\bm\beta(\bm\beta\cdot\bm
E)+\bm\beta\!\times\!\bm B\right)\!\right\}, 
\label{eq7}\end{eqnarray}
where $a=(g-2)/2$. A comparison of Eqs. (\ref{eqspinm}) and (\ref{eq7}) shows that the horizontal projections of vectors $\bm\Omega$ and $\bm\Omega^{(cyl)}$ coincide.

To find the angular velocity of spin motion in the FS
coordinate system, it is necessary to subtract an angular velocity of rotation of the $\bm N$ vector from $\bm\Omega$. This way, in
this reference frame, an angular velocity of spin rotation is found to be \cite{GBMT}:
\begin{eqnarray}
\bm\Omega^{(FS)}=-\frac{e}{m}\left[a\bm B-\frac{a\gamma}{\gamma+1}\bm\beta(\bm\beta\cdot\bm B)+\left(\frac{1}{\gamma^2-1}-a\right)\left(\bm\beta\times\bm
E\right)\right.\nonumber\\
+\left.\frac{\eta}{2}\left({\bm E}-\frac{\gamma}{\gamma+1}\bm\beta(\bm\beta\cdot\bm E)+{\bm\beta}\times {\bm B}\right)\right].
\label{Nelsonh}\end{eqnarray}
Equation (\ref{Nelsonh}) is commonly used in the literature to
describe a spin motion in accelerators and storage
rings.

Of course, Eq. (\ref{Nelsonh}) is more compact. However, this
compactness is achieved owing to the fact that the
deviation angles from the vertical line for the FS coordinate system axes change with time, while the radial
and azimuthal axes of the cylindrical coordinate system always belong to a horizontal plane. Therefore, Eq. (\ref{Nelsonh}) can in particular create the illusion that, under regular
conditions ($\bm\beta\cdot\bm B=0$), the efficiency of impact of vertical and radial fields on a spin is identical. However, when $\bm E=0$ and the EDM is neglected, the ratios $\Omega_\rho^{(cyl)}/B_\rho$ and $\Omega_z^{(cyl)}/B_z$ differ in
fact by $(a\gamma+1)/(a\gamma)$ times.
For leptons (an electron and a muon) this ratio can be
very large. The reason is that, to determine an \emph{observable} effect, the motion of tangential and normal axes
of the FS coordinate system should be added to the spin motion in
this coordinate system. Once this factor is taken
into account, both coordinate systems give an equivalent description of the spin motion.
As an example demonstrating the necessity of correct taking into account a spin rotation around the horizontal coordinate axes, let us mention the experiment
that measured the anomalous magnetic moment of a
muon \cite{PRDfinal}. In this experiment, the muon momenta
satisfied the condition $1/(\gamma^2-1)=a$. The presence of
a weak radial magnetic field and vertical electric one,
which compensates for the impact of the former on the
particle motion, leads to a certain increase in the
angular velocity of spin rotation: $\sqrt{\Omega_z^2+\Omega_\rho^2}$ instead of $\Omega_z$. A naive usage of Eq. (\ref{Nelsonh}) for calculating $\Omega_z$ without
taking into account the motion of tangential and normal axes of the FS coordinate system leads to an
expression that substantially differs from the correct
result given by Eq. (\ref{eq7}). Note that a thorough elimination of magnetic field inhomogeneities in the experiment \cite{PRDfinal} allowed a significant reduction in the radial
magnetic field; its contribution to the final angular
velocity of spin rotation in this latter experiment was
negligible.

\section{CONCLUSIONS}

The comparative analysis of a description of spin
dynamics in the cylindrical and Frenet–Serret coordinate systems carried out in this paper has demonstrated
that it is possible to use both reference frames. The
advantage of the FS coordinate system is a mathematical apparatus thoroughly developed on its basis for
describing spin evolution of particles and nuclei in
accelerators and storage rings. However, the cylindrical coordinate system may as well be efficiently used
for that purpose. Its advantage is the absence of
motion of the coordinate axes relative to the plane of
unperturbed motion of particles/nuclei and, consequently,
with respect to the stationary detectors.

\vskip 3mm
This work was supported by the Belarusian Republican
Foundation for Fundamental Research (Grant No. $\Phi$14D-007).

\end{document}